\documentclass{emulateapj}
\usepackage{natbib,amsmath}
\usepackage{graphicx}
\usepackage{hyperref}

\usepackage{float}
\usepackage{caption}
\usepackage{subcaption}

\providecommand{\e}[1]{\ensuremath{\times 10^{#1}}}

\begin{document}
\title{Stars Get Dizzy After Lunch}

\author{Michael Zhang}
\affil{Department of Astrophysical Sciences, 5491 Frist Center,
Princeton University, Princeton, NJ 08544, USA}
\author{Kaloyan Penev}
\affil{Department of Astrophysical Sciences, 4 Ivy Lane, Peyton Hall,
Princeton University, Princeton, NJ 08544, USA}

\begin{abstract}
Exoplanet searches have discovered a large number of 'hot Jupiters'--high-mass
planets orbiting very close to their parent stars in nearly circular orbits.
A number of these planets are sufficiently massive and close-in to
be significantly affected by tidal dissipation in the parent star, to a degree 
parametrized by the
tidal quality factor $Q_*$. This process speeds up their stars' rotation rate 
while reducing the planets' semimajor axis.  In this paper, we investigate the 
tidal destruction of hot Jupiters.
Because the orbital angular momenta of these planets are a significant fraction
of their stars' rotational angular momenta, they spin up their stars 
significantly while spiralling to their deaths.  Using Monte Carlo simulation, 
we predict that for $Q_* = 10^6$, $3.9\e{-6}$ of stars with the Kepler Target 
Catalog's mass distribution should have a 
rotation period shorter than 1/3 day (8 h) due to accreting a planet.  
Exoplanet surveys such as SuperWASP, HATnet, HATsouth, and KELT have already
produced light curves of millions of stars.  These two facts suggest that it
may be possible to search for tidally-destroyed planets by looking for stars 
with extremely short rotational periods, then looking for remnant planet cores 
around those candidates, anomalies in the metal distribution, or other 
signatures of the recent accretion of the planet.
\end{abstract}
\section{Introduction}
\cite{tidal_evolution_1973} showed that a planet-satellite system evolving
under tidal dissipation has three outcomes: the satellite could spiral inwards
to its death, spiral outwards to escape, or approach a tidally locked
equilibrium.  Ever since the discovery of the first exoplanets, astronomers 
have studied tidal decay due to exoplanet-star interactions.  For example, 
\cite{51pegasi_tides} studied tidal dissipation in 51 Pegasi b--the first
exoplanet discovered around a main-sequence star--and concluded that the
planet's orbit is unstable, even though the decay time is longer than its
star's main-sequence lifetime.  Most of the early exoplanets were similar to 
51 Pegasi b--they are hot Jupiters, with high masses and short orbital periods.

Tidal decay in planetary systems has been relied on to explain a wide variety
of phenomena.  For example, it is hypothesized to be responsible for 
circularizing eccentric orbits--although this is due to tides in the planet, 
not in the star.  The standard theory of hot Jupiter formation suggests
they formed outside the ice line and migrated closer to their stars by one or
both of two mechanisms: interaction with the protoplanetary disk or with
other planets. At this early age, the star's rotation 
frequency would be faster than the planet's orbital frequency and increasing
due to shrinking of the star, so tidal forces would act to push the planet 
outwards 
and prevent it from spiralling into the star (\cite{disk_migration_tides}).
Tidal decay has also been implicated in
the apparent correlation between high obliquity, as measured by the 
Rossiter-McLaughlin effect, and high stellar temperature. The hypothesis is 
that higher-mass (and hotter) stars have much thinner convective zones, and
are thus subject to much less tidal dissipation.  A lower-mass star would have
a significant convective zone, causing a massive planet's orbit to
significantly evolve towards alignment (\cite{hot_stars_obliquity}).  More
recently, \cite{albrecht_obliquities} examined a larger set of RM measurements
and found that the temperature dependence in \cite{hot_stars_obliquity}
still holds.  Additionally, they found that obliquity is smaller for hot
Jupiters where the expected tidal timescale is short, consistent with a
hot Jupiter formation mechanism where the planets' initial obliquity is random.

Depending on the tidal dissipation efficiency, massive hot Jupiters close to
their stars may spiral in towards the Roche limit and be destroyed in a
short timespan, depositing their angular momentum on the star's envelope while
leaving behind a small, rocky, low-period core. \cite{short_period_objs}
carried  out a search for extremely low-period transiting objects in the Kepler
dataset and found 13 candidates, with periods from 3.3 to 10 hours and
estimated masses from 3-200 Earth masses (with large error bars).  The authors
suggest the possibility that these objects are the remnants of destroyed
hot Jupiters.  \cite{teitler_konigl} investigated the phenomenon
discovered by \cite{mma113}--namely, the dearth of Kepler Objects of Interest 
(KOIs) with orbital periods smaller than 2-3 days around stars
with rotation periods shorter than 5-10 days.  They use a Monte Carlo model
to select an initial population of planets and evolve them forward in time,
accounting for tidal interaction, magnetic braking, and core-envelope
coupling.  The authors conclude that the observed distribution of low orbital
period, low rotation period systems is qualitatively consistent with the 
simulated distribution if $Q_*$ is around $10^5$ to $10^6$.

At least two examples of tidal destruction may have already been discovered.
\cite{potential_engulfment} reports the discovery of BD+48740, a star with
unusually high lithium content and with a 1.6 $M_J$ planet in a highly
eccentric ($e \approx 0.67$) orbit.  The authors suggest that both
characteristics--both unusual for an evolved star--can be explained by the 
recent engulfment of a more inner planet.
They also note that current data is not accurate enough to verify this
hypothesis.  \cite{corot7b} examines CoRot-7b, the first confirmed rocky
exoplanet, and suggests that it could have originated as a farther-out gas 
giant, after which evaporation and tidal decay reduced it to its current
state.

\cite{planet_death} found that the destruction of a hot Jupiter might
be observable on a human timescale. The authors predict that destruction 
happens at a galactic rate of $0.1-1 yr^{-1}$, and describe 3 qualitatively 
different scenarios, depending on the planet-to-star density ratio 
$\rho_p/\rho_*$. If $\rho_p/\rho_* > 5$, the planet plunges directly
into the star, creating a EUV/soft x-ray transient that lasts weeks to months
and an optical transient lasting days.  If $\rho_p/\rho_* < 1$, the planet
reaches the Roche radius and stably transfers mass to its star over a timescale
of ~1000 years--the most difficult scenario to observe.  
If $1 < \rho_p/\rho_* < 5$, the planet is disrupted into an
accretion disk, which causes an optical transient that lasts weeks to months.
Optical transients are expected to be similar to, but distinguishable from,
classical novae.

This paper takes a parallel course to \cite{planet_death} and estimates the
fraction of Sun-like stars in the galaxy that have swallowed a planet and
are rotating at extremely short rotation periods ($P < 8h$) as a result. 
Extreme rotation periods
might be good tracers for planet death because few stars, especially on
the main sequence, naturally spin this fast.  Exoplanet
surveys such as SuperWASP (\cite{superwasp}), HATnet (\cite{hatnet}),
HATsouth (\cite{hatsouth}), and KELT (\cite{kelt})
have produced light curves of millions of 
stars to detect planetary transits.  Light curves produced by these surveys are 
ideal for finding short-rotation-period stars.  8h is much shorter than the
orbital periods of the planets that the surveys routinely detect, and
starspots are not likely to change significantly over 8h, giving a very
highly periodic photometric signal.  The question of how many stars amongst 
these millions have detectable rotation periods is addressed in Section
\ref{subsec:detectability}.

To make this estimate, we need to know the tidal dissipation effiency.  
Unfortunately, the tidal dissipation efficiency is very poorly known, 
and proposed values range from $10^6 - 10^{12}$ (\cite{tidal_q_range}).  
Depending on the
value of $Q_*$, the planet could have a significant effect on the star's
rotation rate, or it could have a negligible effect for the entirety of the
star's lifetime.  Data on the circularization of binary stellar orbits seem
to indicate $Q_* \approx 10^5$ or $Q_* \approx 10^6$ 
(\cite{binaries_tidal_circ}).  However, \cite{ogilvie_planet_tides} propose
that hot Jupiters should have smaller dissipation efficiency than binary stars
because the Hough waves they excite are not damped at the center of the star.
\cite{penev_sasselov} also argue that there is good reason to suspect
that $Q_*$ may not be $10^6$ for planet-hosting stars due to differences in
the mechanism of dissipation.  For example, the members of a binary stellar 
system are likely to be tidally locked, whereas most known transiting hot 
Jupiters do not have enough orbital angular momentum to synchronize their 
stars' rotation with their orbit.

Numerous planets have been used to constrain $Q_*$. \cite{hellier_wasp18}
analyzed WASP-18b, the first discovered planet with a period less than 1 day, 
and found
that a $Q_*$ of $10^6$ would mean the planet's remaining lifetime is less than a
thousandth of the star's main-sequence lifetime, and that the shift in transit
timing would be detectable after only 10 years.  It should be noted that
WASP-18 is a 1.24 solar mass star, and that tidal dissipation may be less
efficient in stars of this high mass due to the absence of a convective zone.
\cite{hellier_wasp19} analyze WASP-19b and conclude that likely values for 
$Q_*$ are $10^7-10^8$, but because
the age of WASP-19 is highly uncertain, the best estimate being that it has a
65 percent chance of being older than 1 Gyr with estimates ranging from a few
hundred Myr to many Gyr (\cite{wasp19_age}), no firm conclusions can be drawn.  
\cite{penev_et_al} argue that the distribution of planet orbital periods is 
inconsistent with $Q_* < 10^7$ with 99 percent probability.  Because $Q_*$ is
so uncertain, we will run simulations for $Q_*=10^6$, $Q_*=10^7$, and
$Q_*=10^8$.

This paper is divided into 4 further sections. The first, 'Model', will 
describe the set of ODEs we use to model the evolution of a stellar system, as
well as our choice of parameters for the ODEs.  'Simulation method' will
describe the code used to solve these ODEs, the planets chosen for constraining
Q, and the distributions of planetary and stellar parameters used to estimate
the number of fast rotators.  'Results' has a self-evident name.
'Conclusions' will discuss the possibility of detecting planets as they 
inspiral into their stars and shortly afterwards.

\section{Model}
Our model is the same as the one used by \cite{penev_et_al}.  Planets are
assumed to be in circular orbits and tidally locked to their stars.  Tidal
dissipation within the planet itself is therefore negligible.
The dynamical evolution of a stellar system is assumed to be governed by the
following differential equations:
\begin{align}
\frac{da}{dt} &= sign(\omega_c - \omega_{orb})\frac{9}{2}\sqrt{\frac{G}{aM_*}}(\frac{R_*}{a})^5\frac{m_p}{Q_*}\\
(\frac{dL_c}{dt})_{tide} &= -\frac{1}{2}m_pM_*\sqrt{\frac{G}{a(M_*+m_p)}}\frac{da}{dt}\\
(\frac{dL_c}{dt})_{wind} &= -K\omega_cmin(\omega_c, \omega_{sat})^2(\frac{R_*}{R_{\Sun}})^{1/2}(\frac{M_*}{M_{\Sun}})^{-1/2}\\
\frac{dL_c}{dt} &= \frac{\Delta L}{\tau_c} - \frac{2}{3}R_r^2\omega_c\frac{dM_r}{dt} + (\frac{dL}{dt})_{wind} + (\frac{dL}{dt})_{tide}\\
\frac{dL_r}{dt} &= -\frac{\Delta L}{\tau_c} + \frac{2}{3}R_r^2\omega_c\frac{dM_r}{dt}\\
\Delta L &= \frac{I_cL_r - I_rL_c}{I_c+I_r}
\end{align}

where a is the planet's semimajor axis, $M_*$ is the star's mass, $m_p$ is the
planet's mass, $\omega_{orb}$ is the planet's orbital frequency, 
$L_c,I_c,\omega_c$ are the convective zone's angular momentum,
moment of inertia, and rotation rate respectively, $I_r,I_r,\omega_r,M_r,R_r$ 
are the radiative zone's angular momentum, moment of inertia, rotation rate,
mass, and radius respectively,
and $K, \omega_{sat},\tau_c$ are constants set by the rotation model.  The 
rotation model will be discussed in more detail in the next subsection.

Equation 1 is given by \cite{Goldreich_63, Kaula_68, Jackson_et_al_08a}.
Equation 2 follows from conservation of momentum, assuming that momentum lost
by the planet is added to the star's convective zone.  Equation 3 is taken from
\cite{Stauffer_Hartmann_87, Kawaler_88, Barnes_Sofia_96}, and models the
magnetic spindown of the star.  Equation 5 uses a formulation by
\cite{MacGregor_91} and \cite{Allain_98} to model the interaction between
core and envelope, while Equation 4 follows trivially from Equation 5. 

The quantities $I_c, I_r, M_r, R_r$ and $R_*$ are taken from 
stellar evolution tracks computed using the YREC model (\cite{yrec}), 
which contain these quantities at various ages for masses of
0.5, 0.6, 0.7, 0.8, 0.9, 1.0, 1.05, 1.1, 1.15 and 1.2 $M_{\Sun}$. Cubic
spline interpolation was used to interpolate between the ages for each mass,
after which another cubic spline interpolation was used to interpolate between
masses.  Since this interpolation is only valid for 0.5-1.2 $M_{\Sun}$, we
never simulate stars outside this range anywhere in the paper.  
Because high-mass
stars have thin convective zones, which may cause different physical mechanisms
to dominate the tidal dissipation and lead to a different $Q_*$, we further 
restrict ourselves to
$M_* < 1.05M_{\Sun}$ throughout the paper.

We assume that the planet forms at 5 Myr, and that its orbit does not
change by any mechanism except tidal dissipation in the star.  
This formation scenario is
justified if the planet is brought to its initial orbit by disk interaction.  
It might not be justified if the planet arrives via Kozai resonance or 
planet-planet scattering, followed by tidal circularization due to tides raised
on the planet.

\subsection{Rotation model}
Equation 3 above models angular momentum loss due to the stellar magnetic
wind.  This model, the same as the one used by \cite{irwin_bouvier}, reproduces
Skumanich's law until $\omega_{conv} \geq \omega_{sat}$, at which point the rate
of angular momentum loss becomes proportional to the rotation rate as opposed
to its cube.  In addition to $\omega_{sat}$, a planetless star's rotational
evolution depends on two parameters: the magnetic wind strength K, and the
core-envelope coupling timescale $\tau_c$.

This paper focuses its attention on stars of 0.5-1.05 solar masses.  Stars of
this mass initially have a wide distribution of rotation rates. 
\cite{irwin_bouvier}
compiled 3100 rotation periods of stars in open clusters and found rotation
periods ranging from less than 0.1 days to more than 10 days in clusters
younger than 5 Myr.  This paper, just like 
Irwin \& Bouvier, will assume the convective rotation rate is held constant 
before the
disk dissipates. 

\begin{figure}[t]
\includegraphics [width= 0.45\textwidth]{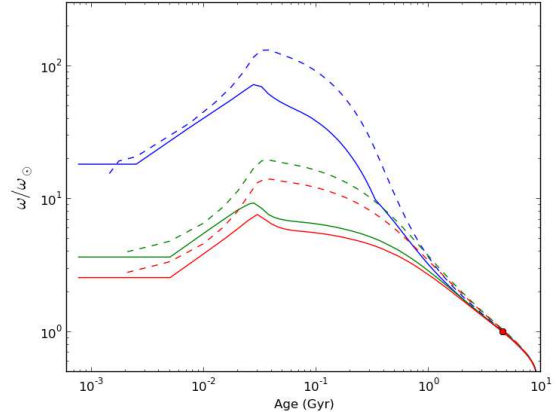}
\caption{Rotation speeds of the core (dashed lines) and envelope (solid lines) 
for 3 rotation scenarios, as simulated by our code.  Compare with Figure 3 from
\cite{gallet_bouvier}.}
\label{fig:rotation_evolution}
\end{figure}

We take our values for $\omega_{sat}$ and $\tau_c$ from \cite{gallet_bouvier}.  
They use a
different rotation model, but it turns out that our model approximates their
model decently well, as shown in Figure~\ref{fig:rotation_evolution}.  
Furthermore, they derived their fits by eye, so we do not lose a substantial 
amount of accuracy.

The authors quote a single value for $\omega_{sat}$, namely 10 times the solar
rotational frequency, or 2.45 rad/day. 
They fit three prescriptions for $\tau_c$: one for the fast
rotators, one for the intermediate rotators, and one for the slow rotators.
A fast, intermediate, and slow rotator have initial rotation periods
of 1.4, 7, and 10 days respectively.  The circumstellar disk is assumed to
dissipate at 2.5, 5, and 5 Myr respectively.
The intermediate and slow rotators have similar evolutions, and similar
parameters.  For each prescription, we found the K that reproduces the Sun's
rotation rate at its current age of 4.57 Gyr.  All 3 prescriptions are shown
in Figure \ref{fig:rotation_evolution}, and their parameters are summarized in
Table \ref{table:rot_params}.

\begin{table}[H]
\begin{center}
$
\begin{array}{lllllr}
    \hline
\text{Prescription} & \text{Init} P_{rot} & K & \tau_c  & 
    \omega_{sat} & T_{disk}\\ 
\hline
Fast &1.4 &0.155 &12 &2.45 & 2.5\\
Intermediate &7 &0.17 &28 &2.45 & 5\\
Slow &10 &0.17 &30 &2.45 & 5\\
\hline
\end{array}
$
\end{center}
\caption{Rotation parameters for 3 different prescriptions. Units are days
for $P_{rot}$, Myr for $\tau_c$, and Myr for $T_{disk}$.
}
\label{table:rot_params}
\end{table}

\section{Simulation method}
The simulation code consists of 3 main components:
1. the orbit solver
2. Initial condition solver
3. Fast-rotator estimator

\subsection{Orbit solver}
This code is a revamped and much-improved version of the code written by 
\cite{penev_et_al}.  The code is described by \cite{code_article}.

The orbit solver uses the GNU Scientific Library (GSL) to solve the system of
ODEs with respect
to time.  At the beginning of a star's life, when there is no planet and the
radiative zone has yet to form, the convective rotation rate is locked to that
of the disk and there is no need for simulation.  After the
radiative zone forms, we have 2 variables: $L_{conv}, L_{rad}$.  After the
planet also forms, we have 3 variables: $L_{conv}, L_{rad}, a$.  When the planet
dies, we again have only 2 variables.  The orbit solver automatically detects
which variables are relevant at each timestep, and uses the correct set of
differential equations to advance to the next timestep.

To improve both speed and stability, we use GSL's implicit Bulirsch-Stoer
stepper function.

We used $a^{6.5}$ as the dependent variable instead of a, because at the end of
the planet's lifetime, a decreases extremely rapidly with time.  This creates
accuracy and performance issues with GSL's ODE solver.  $a^{6.5}$
behaves more nicely because $\frac{da^{6.5}}{dt}$ does not exhibit the same
sensitive dependence on a as $\frac{da}{dt}$.  

\begin{figure}[h]
\includegraphics [width= 0.45\textwidth]{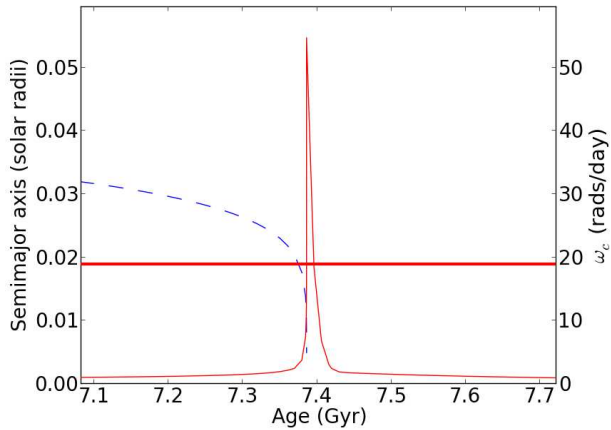}
\caption{HAT-P-20b spiralling into its star.  The blue dashed line represents 
the semimajor axis; the red line represents the star's rotation rate.  The bold
red line is the rotation threshold we use, $P_{rot}=8h$.}
\label{fig:planet_inspiral}
\end{figure}

For illustrative purposes, Figure \ref{fig:planet_inspiral} shows the evolution
of semimajor axis and spin rate for one particular stellar system, namely
HAT-P-20, as simulated by our code.  Notice the huge
jump in spin rate after the star swallows its planet, and the rapid decline
in spin rate afterwards.

\subsection{Fast rotator estimator}
This portion of the code is run for 3 assumptions about $Q_*$ 
($10^6, 10^7, 10^8$).

The code selects a star of random mass and random initial rotational period, 
where 'initial' means at 5 Myr.  It selects a planet with random
initial semimajor axis, random radius, and random mass, and puts it in orbit
around the star.  The orbit solver is then used to evolve the system until the
planet passes through the Roche radius, the planet passes through the star's
radius, the star dies, or the maximum age of 10 Gyr is reached, whichever comes
first.  10 Gyr is the limit of the YREC stellar models that we use, but since
it is also close to the age of the universe, it prevents low-mass stars from
living longer than the age of the universe.
The Roche radius r and the lifetime l are computed 
using:
\begin{align*}
l &= (9 Gyr)(\frac{M_*}{M_{\Sun}})^{-3}\\
r &= 2.44r_p(\frac{M_*}{m_p})^{1/3}
\end{align*}

Note that this definition of lifetime is somewhat contrived, since a more
accurate expression is $(10 Gyr)(\frac{M_*}{M_{\Sun}})^{-2.5}$.  It was chosen
purely due to limitations in the YREC tracks, but for masses around 1 solar
mass, it is close to the 'actual' main-sequence lifetime.  We don't consider
masses above 1.05$M_{\Sun}$, and low-mass stars have lifetimes above 10 Gyr, so
their precise lifetimes are never used.

If the planet is swallowed by the star, which happens if the planet passes the 
Roche
radius or the stellar surface, the orbit solver removes the planet, adds its
leftover angular momentum to the star's convective zone, and
continues evolving until l or 10 Gyr is reached.  It then computes the total 
amount of
time during which the star completes more than 3 revolutions per day.  
This threshold was chosen so that for all stars in this simulation, the natural
evolution of the star (without any tidal effects from the planet) would not
cause its rotation speed to exceed the threshold at any point.
We assume that the age distribution of stars is uniform, meaning that
 the
probability P of catching this specific star rotating very fast is proportional
to the time it spent rotating very fast, divided by min(l, 10 Gyr).

We assume that the planet is destroyed instantaneously upon touching the star
or passing the Roche radius.  \cite{planet_death} describes the process of 
planet destruction
and lists three qualitatively different scenarios, depending on the density
ratio $\rho_p/\rho_*$.  If $1 < \rho_p/\rho_* < 5$, the planet reaches the
Roche radius and stably transfers mass to the star over a timescale of
thousands of years.  If the density ratio is lower, the planet still reaches
the Roche radius, but quickly breaks up and accretes onto the star on a
timescale of days to weeks.  If the density ratio is higher, the planet plunges
into the stellar atmosphere and is destroyed in about 100 days.  Since even
1000 years is negligible compared to the megayear timescales that fast 
rotation rates are expected to last, our approximation of instantaneous
accretion is very safe.

This process of picking a random star and random planet is repeated until
10,000 stars exceed the threshold of $P_{rot} = 8h$ due to the
death of their planets. Intermediate results are printed out to make sure the 
results converge.

Every aspect of this random sampling process will now be discussed.

\subsubsection{Stellar mass distribution}
The mass distribution is taken from the Kepler Target Catalog, which are the
stars in the Kepler Input Catalog selected for examination by Kepler.  We
take the mass distribution from Kepler because it is a convenient catalog of
stellar information, and because it is meant to catalogue Sun-like stars--not
too dissimilar to the stars present and future transit surveys focus on.\\

Stars with a mass less than 0.5 or more than 1.05 $M_{\Sun}$ were excluded.  
A bin is randomly selected according to a histogram representing the
distribution.  The stellar mass is then chosen uniformly at random between the
two bin edges.  For example, if the bin (0.5, 0.506) is chosen, the mass is
chosen from the range 0.5-0.506 with uniform probability.

\subsubsection{Planetary orbit distribution}
\cite{kepler_occurrence} uses Kepler data to derive the current distribution of 
planets with respect to their period P and their radius R.  
The distributions they found were:
\begin{align*}
\frac{df}{dlogP} &\propto P^{\beta}(1-e^{-\gamma \frac{P}{P_0}})\\
\frac{df}{dlogR} &\propto R^{\alpha}
\end{align*}

$\frac{df}{dlogP}$ is the fraction of planets in a logarithmic interval 
centered on P.
$\alpha = -1.92$.  $\beta$,
$\gamma$, and $P_0$ depend on radius, but we only consider planets with radii
greater than 8 $R_{\Earth}$.  This is because planets significantly below this
radius are not hot Jupiters, and are unlikely to have the mass necessary to
significantly influence their stars' rotation rates.  For this radius range
(8-32 $R_{\Earth}$), $\beta=0.37 \pm 0.35$, $P_0 = 1.7 \pm 0.7$, and 
$\gamma = 4.1 \pm 2.5$.  
Additionally, we also only consider planets with a period less than 5.9 days,
under the assumption that planets with longer periods will not inspiral into
their stars.

Note that these distributions are for currently-observed planets, whereas we
want the period distribution when the planets first formed.  To get the latter
from the former, we first assume that the initial distribution be represented
with the same 3 parameters: $\beta, P_0, \gamma$.  We created an algorithm 
which takes these 3 initial parameters, randomly samples planets from that 
distribution, and evolves them for a random amount of time.  The resulting 
distribution is compared to the modern-day distribution using the KS 
test--the lower the KS statistic, the better the fit.  A simulated annealing 
algorithm was used to sample the 3-dimensional parameter space and find the
best initial distribution for every Q.  The results are presented in
Table \ref{table:init_period_dist}.

\begin{table}[H]
\begin{center}
$
\begin{array}{lllr}
    \hline
Q_* & \beta & P_0 & \gamma \\
\hline
10^6 & -1.4 & 1.1 & 16.3\\
10^7 & 0.088 & 1.5 & 16.2\\
10^8 & 0.44 & 1.6 & 4.1\\
\hline
\end{array}
$
\end{center}
\caption{Initial orbital period distributions}
\label{table:init_period_dist}
\end{table}

\subsubsection{Planetary mass distribution}
We assume that the planetary mass distribution is independent of radius.  For
hot Jupiters ($R > 8R_{\Earth}$), this is roughly true, as shown by
exoplanets.org data and well-supported by theory.

The mass distribution we use was taken from exoplanets.org. Only planets 
discovered by transit searches with a radius greater than 8 $R_{\Earth}$, a
period smaller than 5.9 days, and a known mass were included.  For hot
Jupiters, detection probability is not highly correlated with radius.  It is
however highly correlated with orbital period--even for periods as short as 
5.9 days, longer-period planets are much less likely to be detected.
However, since we are making the assumption that orbital period is not highly
correlated with radius or mass for this radius range--the same assumption that
\cite{kepler_occurrence} make due to lack of data--there is no need to
correct for observational biases.

A bin is randomly selected according to the distribution in the
histogram.  The planet mass is then chosen uniformly at random between the two
bin edges.  The algorithm is exactly the same as the one used for selecting
stellar masses.

\subsubsection{Stellar rotation distribution}

This distribution is taken from \cite{irwin_cluster_data} for the open 
cluster NGC 2362, which is 5 Myr old.  

The data is histogrammed into 10 bins. A random rotational period is generated 
by taking the star's mass, finding its corresponding bin, randomly choosing one 
of the stars in the bin, and returning its rotation rate.

We always choose K and $\tau_c$ based on this randomly-chosen rotation rate.
Whichever prescription the star's initial rotation rate is closest to in 
Table \ref{table:rot_params}, the corresponding parameters are chosen and used
throughout the evolution.

\section{Results and Discussion}

According to \cite{fressin_2013}, the rate of hot Jupiters per star is
0.00252.  With this number, we get the results in 
Table \ref{table:fastrot_results}.  
\begin{table}[H]
\begin{center}
$
\begin{array}{lllllr}
    \hline
Q_* & \text{Stellar fraction} & \text{Sensitivity} & \text{Destruction rate}\\
\hline
10^6 & 2.9\e{-6} & 0.21 & 0.92\\
10^7 & 0.35\e{-6} & 0.52 & 0.37\\
10^8 & 0.08\e{-6} & 0.81 & 0.083\\
\hline
\end{array}
$
\end{center}
\caption{Results of 3 simulations, for 3 different values
of $Q_*$, each of which ran until 10,000 modern-day planets were simulated.
Each simulation was checked for convergence, and all converged long before the
end of the simulation. Second column: fraction of stars that should be
rotating faster than $P_{rot} < 8h$; third column: fraction of planet deaths
that cause their stars' rotation rates to exceed the threshold; final column:
fraction of initial hot Jupiters that are subsequently destroyed}
\label{table:fastrot_results}
\end{table}

\begin{figure*}[t]
\begin{subfigure}[l]{.31\linewidth}
\includegraphics [width= \textwidth]{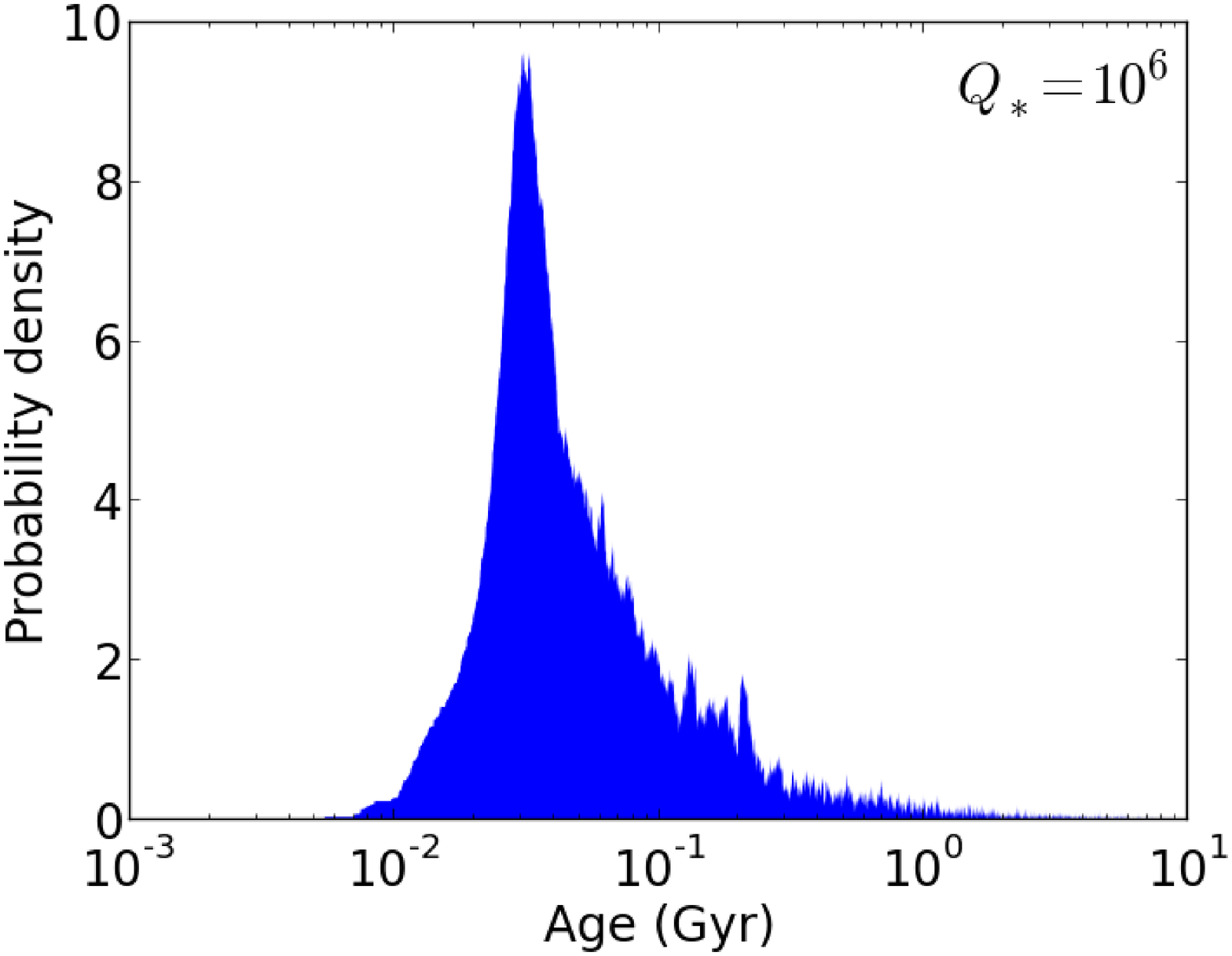}
\end{subfigure}
\begin{subfigure}[c]{.31\linewidth}
\includegraphics [width= \textwidth]{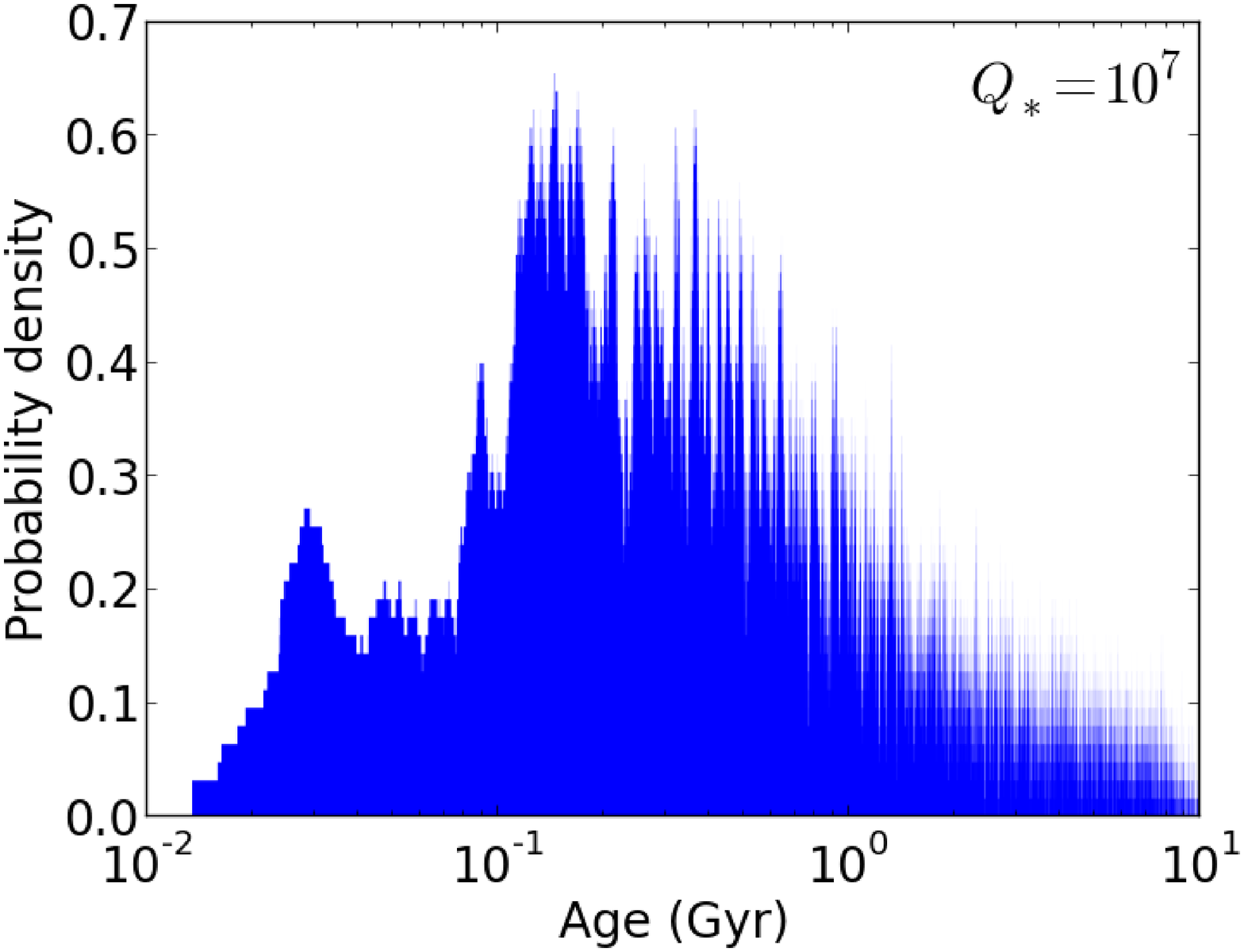}
\end{subfigure}
\begin{subfigure}[r]{.31\linewidth}
\includegraphics [width= \textwidth]{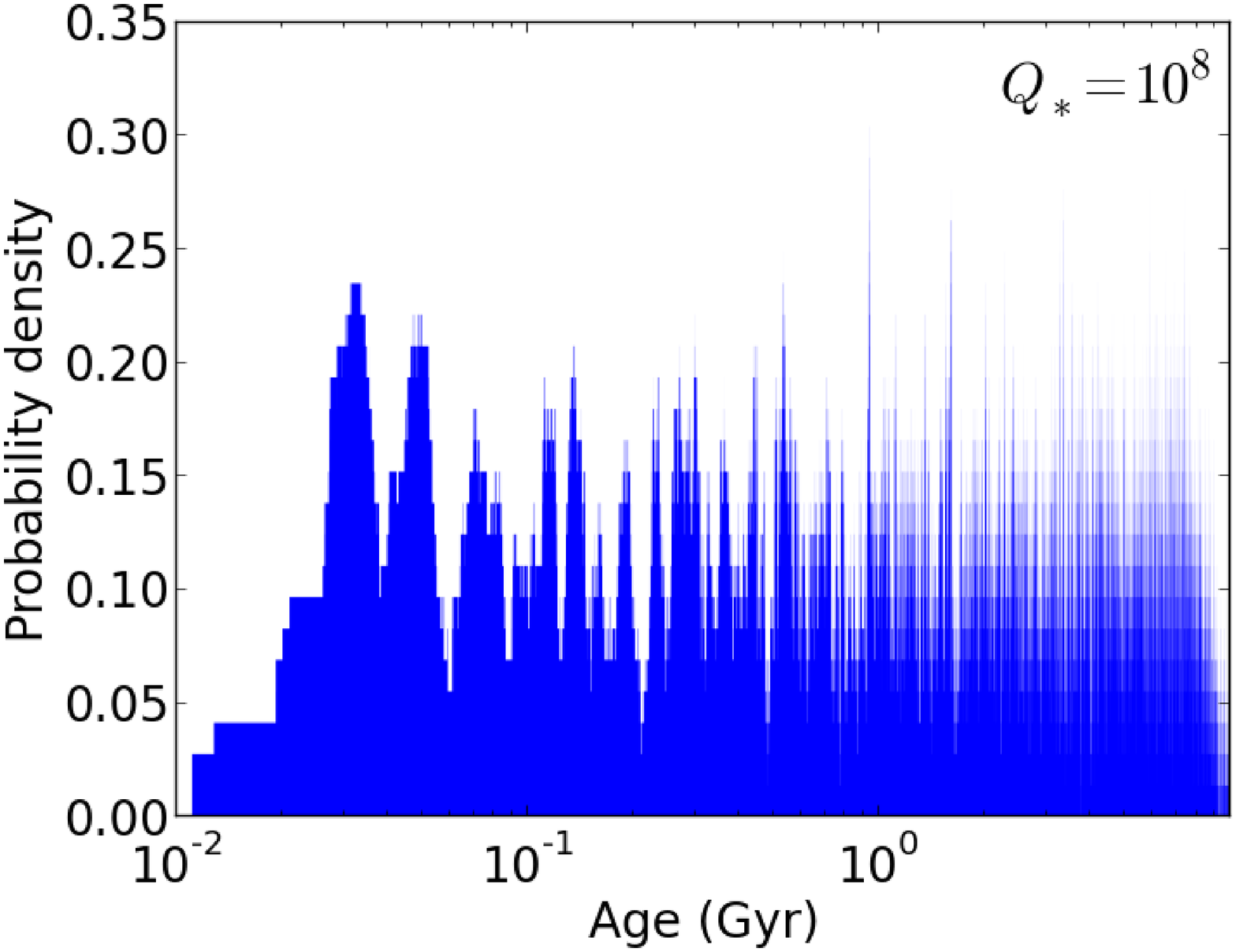}
\end{subfigure}
\caption{Distribution of stellar ages at which the star's spin rate exceeded
threshold due to planet death.}
\label{fig:age_dist}
\end{figure*}

\begin{figure*}[t]
\begin{subfigure}[l]{.31\linewidth}
\includegraphics [width= \textwidth]{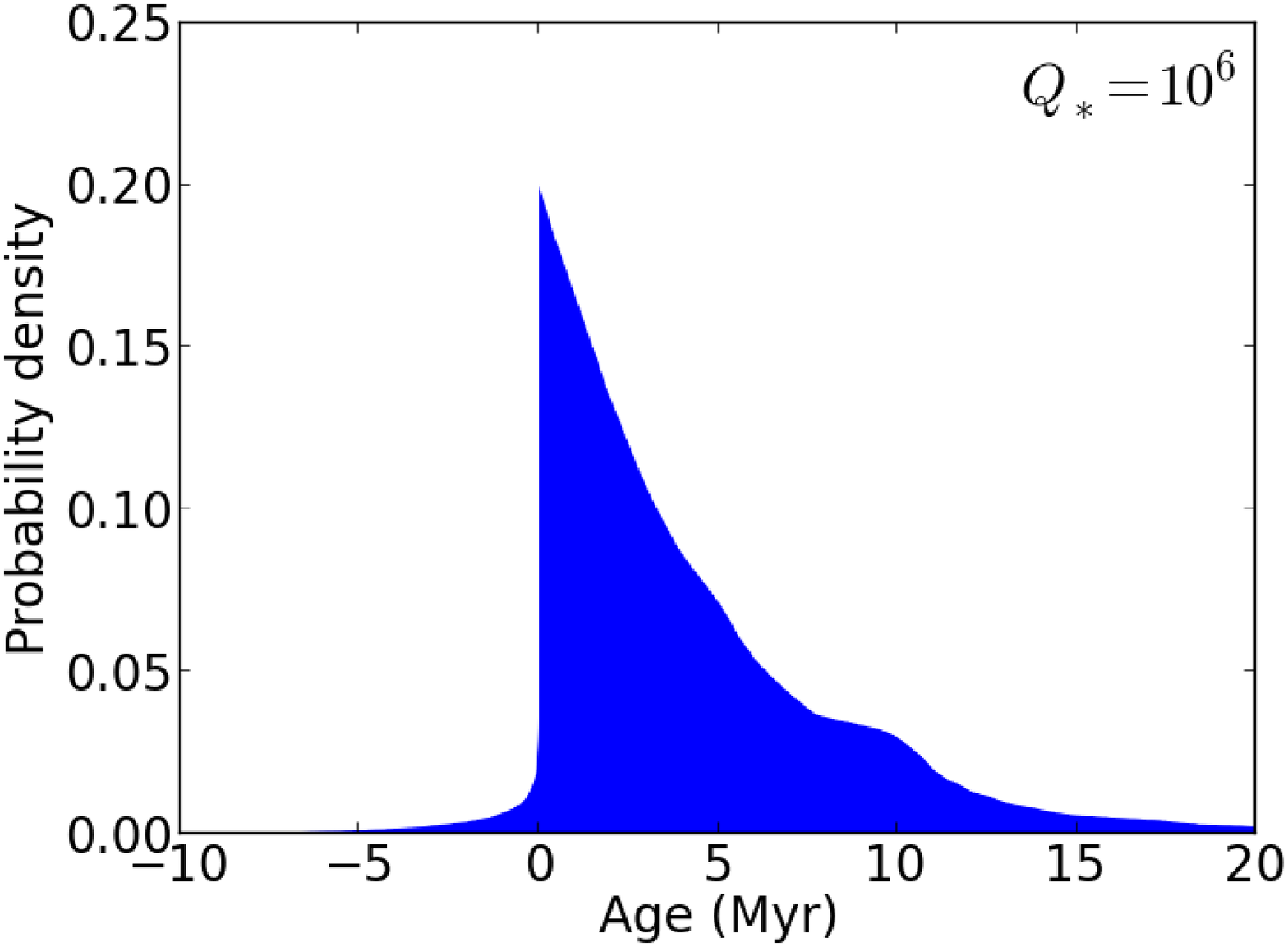}
\end{subfigure}
\begin{subfigure}[c]{.31\linewidth}
\includegraphics [width= \textwidth]{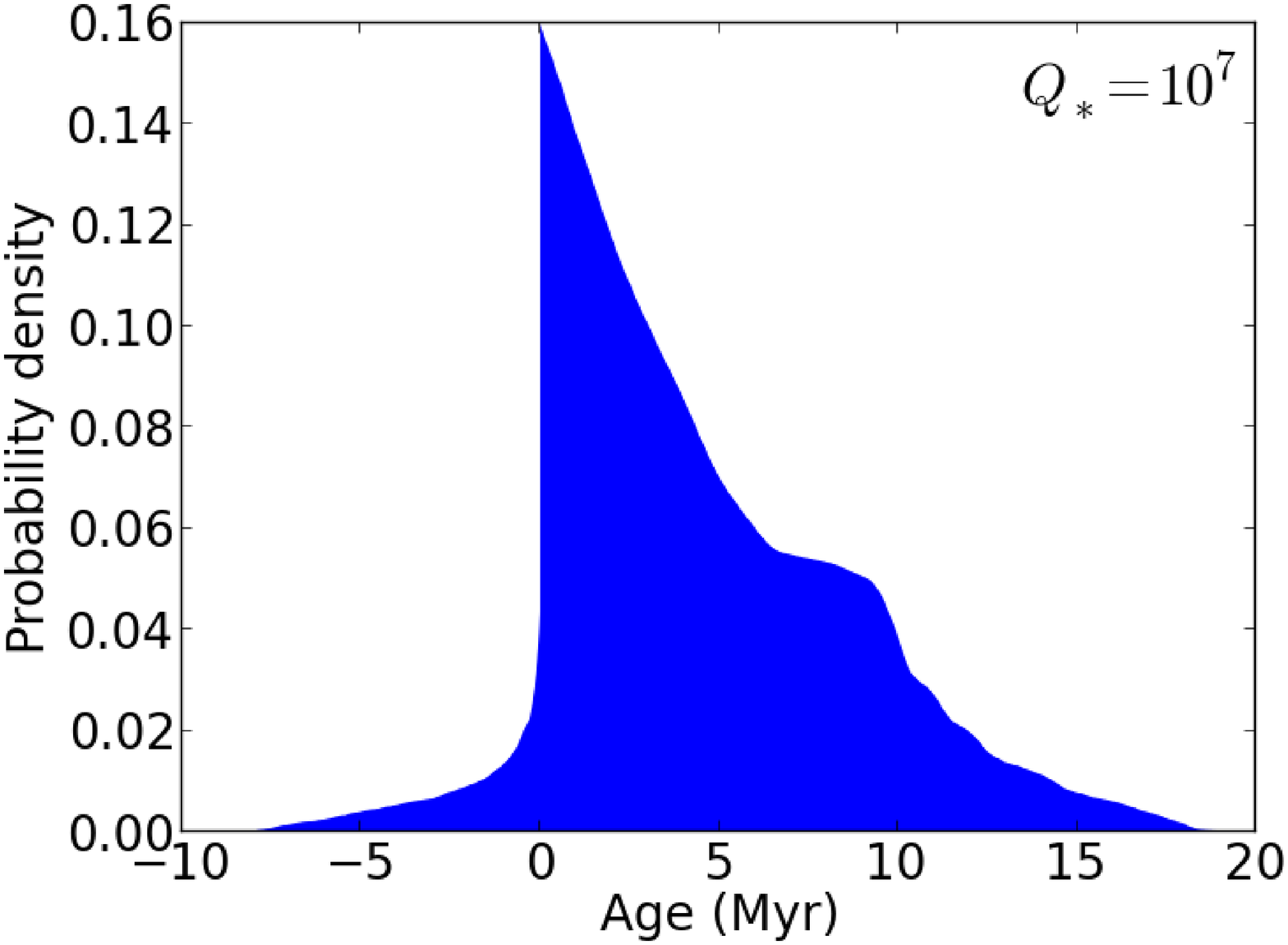}
\end{subfigure}
\begin{subfigure}[r]{.31\linewidth}
\includegraphics [width= \textwidth]{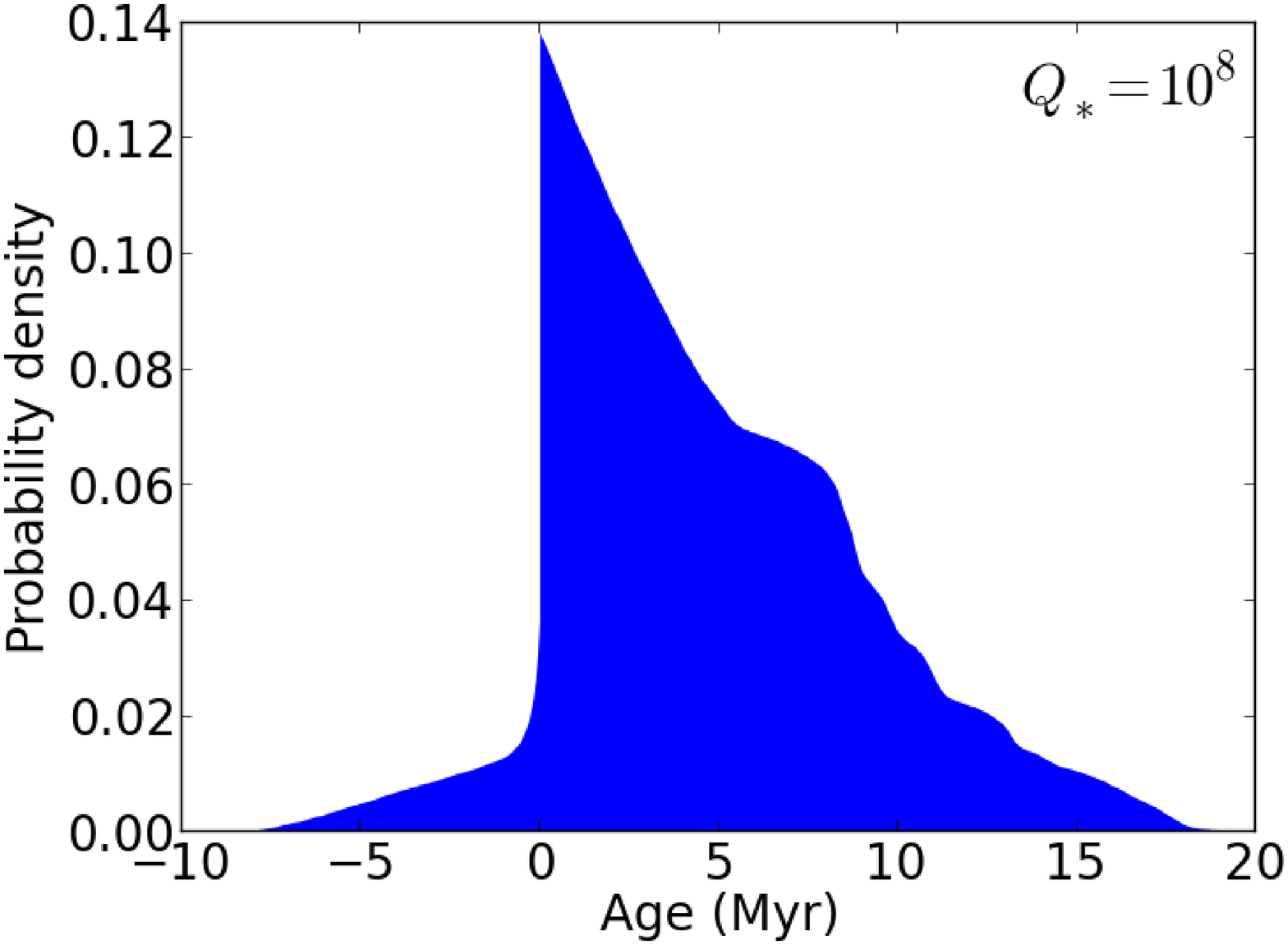}
\end{subfigure}
\caption{These plots represent the fraction of threshold-exceeding stars that
still exceed the rotation threshold ($P < 8h$) at a certain age after planet 
death.}
\label{fig:dur_dist}
\end{figure*}

In Figure \ref{fig:age_dist} we show the distribution of stellar ages at which
the star spins faster than 8h due to planetary death.  Notice that for higher
$Q_*$, the distribution is broader and shifted towards higher ages.  This is
because the initial orbital period distribution has fewer short-period planets 
for high $Q_*$, whereas it exhibits a strong peak at low periods for small 
$Q_*$.  Also notice the spikes at large ages.  These are artifacts, due to the 
small length of time for which stars rotate quickly after planet death.
Because the graphs were constructed by adding up contributions from every dying 
planet, there are some stellar ages that few stars happened to hit when they 
were spinning quickly.

In Figure \ref{fig:dur_dist} we show the distribution of times after planet
death at which the star is spinning quickly.  For example, one can tell that
most stars are still spinning quickly 5 Myr after planet death, whereas 
after 20 Myr, almost all stars have spun down.  Notice that a small but
non-zero percentage of values are negative.  This is because some planets spin
up their stars before dying, while most only spin up their stars after dying
and depositing their angular momentum onto the star's surface.  Notice also 
that the distribution is broader for higher $Q_*$.  This is because a planet 
that formed at higher
semimajor axis has more angular momentum, and therefore deposits more angular
momentum onto the star.  Since the initial period distribution has more
short-period planets for low $Q_*$, the average dying planet deposits less
angular momentum, thereby creating a narrower peak in Figure 
\ref{fig:dur_dist}.

\subsection{Sensitivity of results}
The modern period distribution for exoplanets is highly uncertain.  We have
assumed the period distribution given in \cite{kepler_occurrence}, which has
3 parameters: $\beta, P_0, \gamma$.  Their values and uncertainties are
$\beta=0.37 \pm 0.35$, $P_0 = 1.7 \pm 0.7$, and $\gamma = 4.1 \pm 2.5$.  To
test the sensitivity of our results, we have run our simulation assuming 6
different combinations of these 3 parameters, meant to explore the extremes
of every parameter.  The combinations are given in Table 
\ref{table:orbital_params_sensitivity}.

\begin{table}
$
\begin{array}{lllr}
Index & \beta & P_0 & \gamma\\
\hline
0 & 0.37 & 1.7 & 4.1\\
1 & 0.02 & 1.7 & 4.1\\
2 & 0.72 & 1.7 & 4.1\\
3 & 0.37 & 1.0 & 4.1\\
4 & 0.37 & 2.4 & 4.1\\
5 & 0.37 & 1.7 & 1.6\\
6 & 0.37 & 1.7 & 6.5
\end{array}
$
\caption{Modern orbital period distributions used to test sensitivity. 0 is
nominal.}
\label{table:orbital_params_sensitivity}
\end{table}

For each of the six modern orbital period distributions and each of the three
$Q_*$, we found the initial distribution and ran a simulation to find the
fraction of stars expected to be rotating quickly.  The results are shown in
Table \ref{table:results_sensitivity}.  Notice that the differences between
assumed modern period distributions, for every $Q_*$, are within an order of
magnitude--significantly below the uncertainty in $Q_*$.

\begin{table}
$
\begin{array}{lllr}
\text{Index} & Q_* & \text{Stellar fraction} \times 10^6\\
\hline
0 & 10^6 & 2.9\\
0 & 10^7 & 0.35\\
0 & 10^8 & 0.08\\
1 & 10^6 & 10\\
1 & 10^7 & 0.61\\
1 & 10^8 & 0.106\\
2 & 10^6 & 3.60\\
2 & 10^7 & 0.346\\
2 & 10^8 & 0.067\\
3 & 10^6 & 12.7*\\
3 & 10^7 & 2.14*\\
3 & 10^8 & 0.36\\
4 & 10^6 & 1.67\\
4 & 10^7 & 0.26\\
4 & 10^8 & 0.047\\
5 & 10^6 & 15*\\
5 & 10^7 & 1.07*\\
5 & 10^8 & 0.3\\
6 & 10^6 & 4.6\\
6 & 10^7 & 0.32\\
6 & 10^8 & 0.068\\
\hline
\end{array}
$
\caption{Stellar fraction of fast rotators for every modern orbital period
distribution and every $Q_*$.  Results marked with * are unreliable because no
initial period distribution could be found that closes matches the modern
distribution. In such cases, the closest match was used to compute the result.
0 is the nominal scenario.}
\label{table:results_sensitivity}
\end{table}

\subsection{Detectability of fast rotators}
\label{subsec:detectability}
Recently, \cite{kepler_rot_periods} used an autocorrelation algorithm on
the light curves of 133,030 main-sequence Kepler stars and detected rotation
periods for 34,030 (25.6\%).  The detected rotation periods ranged from 0.2 to
70 days--the former showing that such extreme rotation periods are possible and
detectable.  The authors report a typical range of rotation-induced amplitudes
of 950-22,700 ppm, with a median of 5600 ppm.  The amplitude is highly 
dependent on both period and
temperature, with cool quickly-rotating stars having a much larger amplitude 
than hot quickly-rotating stars.  For comparison, HATNet (\cite{hatnet}) is 
capable of
2 mmag (1800 ppm) photometry on bright stars, with 6 mmag (5500 ppm) being a 
more typical precision.
Thus, HATnet and similar exoplanet surveys should be capable of measuring 
rotation periods for roughly 13\% of the nearby analogues of Kepler 
main sequence stars.

This 25.6\% detection rate does not pose as much of a problem to detecting
planet-swallowing stars with $P < 8h$ as might be expected.  Survey biases 
heavily favor detecting low rotation periods.  A low rotation period would mean
that a given observation timeframe spans many more periods, giving a stronger
signature and reducing the effects of systematic biases that vary with time.
Starspots and other features also change much less from one period to
the next if the period is 8 hours instead of weeks, giving a more strongly
periodic signal.

It should be noted that the detection rate in \cite{kepler_rot_periods} 
depended heavily on temperature.  When they divided stars into
temperature bins, they found
that the lowest temperature bin ($< 4000 K$) had a 83\% detection rate.  They
suggested that misalignment between the star's equator and our line of sight
could account for some of the remaining 17\%.  Their next-lowest bin
(4000-4500 K) has a detection rate of 69\%, and 4000K
corresponds roughly to the 0.5 solar masses that is the lower mass limit of
our study.

It should also be noted that many authors have already used light curves from
ground-based transit surveys to measure stellar rotation rates in clusters.  
For example,
\cite{cameron_coma_rots} reports rotation rates in the 600 Myr Coma Berenices 
open cluster using SuperWASP data, while \cite{delorme_hyades_praesepe_rots}
similarly uses SuperWASP to measure two other clusters of similar age: Hyades
and Praesepe. \cite{hatnet_pleiades_rots} examined
Pleiades members between 0.4 and 1.3 solar masses and detected rotation periods
for 74 percent of stars within the field of view.  This detection rate
increased to 93 percent for stars between 0.7 and 1.0 solar masses.  Of course,
it is easier to detect rotation in cluster stars than in field stars because
young stars have shorter rotation periods and are more variable.  Nevertheless,
fast rotators--aside from being easier to detect--are expected to exhibit
higher-amplitude variations due to higher magnetic activity.  This is shown 
observationally 
(\cite{kepler_rot_periods}, \cite{noyes_obs}) and expected from stellar dynamo 
theory (\cite{earliest_dynamo}) since faster 
rotation produces a stronger magnetic field, which in turn results in higher 
stellar activity.  For example, \cite{diff_rotation} demonstrate
quantitatively, by extending the solar flux transport dynamo model to other 
stars, that the toroidal flux amplitude increases with rotation.
They note that this explains the general trend of Ca II H/K and X-ray 
observations, which are signatures of magnetic activity.

A final concern is that even though the simulation predicts very fast
rotation rates, such rates may exceed the breakup speed of the star and hence
be unphysical.  We do not take this into consideration in our simulation, as a 
detailed examination of the physics of extreme stellar rotation is outside the 
scope of this paper. However, we note that there is no possibility of planetary 
accretion causing the star's envelope to reach escape velocity--after all,
the planet itself is in a tight orbit when it breaks up.  Qualitatively, we
expect the stellar envelope to expand and become more oblate, increasing its 
moment of inertia and decreasing its rotation speed until it no longer exceeds
the breakup speed.  We also note that the breakup period of the Sun is 2.8
hours--significantly below the 8-hour limit used in this paper--and that the
majority of stars in our simulation never reach the breakup period.

\section{Conclusions}

We have simulated the tidal decay of hot Jupiters for 3 different values of
$Q_*$: $10^6, 10^7, 10^8$. We find that, within an order of magnitude, one out
of a million Sun-like stars are expected to exhibit rotation periods less than
8h as a result of swallowing a planet.  Ground-based exoplanet surveys
have examined millions of stars, but only a fraction may exhibit enough
variability to find even a very fast rotation period.  Thus, we conclude that
it is possible for these surveys to detect a fast rotator if $Q_*=10^6$, but 
unlikely if $Q_*=10^8$ or greater.

A very fast
rotation rate is a good indicator because 20-80 percent of hot Jupiter deaths
(depending on $Q_*$) cause their host stars to spin faster than 
$P_{rot} < 8h$, and because there are very few other explanations of such a
high rotation rate:\\
1. The star is young (pre-main-sequence or near ZAMS)\\
2. Tidal interaction in a binary system\\

Scenario 2 can usually be excluded because huge Doppler shifts,
anomalous spectra, and/or transits are all easily detectable features of
binary systems, where 'easily detectable' is with respect to the ease of
detecting a planetary transit.  Scenario 1 is harder to exclude because it is
extremely hard to date stars.  

Other observations may enable confirmation that a planet was tidally
destroyed.  Observing the rocky core is the most direct method, providing that
the planet was not destroyed through a direct-impact merger and that the tidal
disruption process was not violent enough to destroy the core.  If the core is
not detectable, there may be detectable changes in stellar metallicity.
\cite{metallicity} analyzed the change in stellar metallicity due to accretion
of protoplanetary disk material, and found that accretion of a Jupiter-mass
planet can cause changes in the relative metal abundances in the star's 
convective zone that last on the order of tens of Myr.  It is not clear whether
these changes are practically observable, especially given the broadening of
spectral lines caused by the extreme rotation.

\newpage
\bibliography{michael_bib}
\bibliographystyle{apj}

\end{document}